# Two-dimensional Functional Minerals for Sustainable Optics

Ziyang Huang[1], Tianshu Lan[1], Lixin Dai[1], Xueting Zhao[2,3], Zhongyue Wang[1], Zehao Zhang[1], Bing Li[2,3], Jialiang Li[4], Jingao Liu[4], Baofu Ding[1,5]*, Andre K. Geim[1,6], Hui-Ming Cheng[1,2,5,7], and Bilu Liu[1]*

[1] Shenzhen Geim Graphene Center, Tsinghua-Berkeley Shenzhen Institute & Institute of Materials Research, Shenzhen International Graduate School, Tsinghua University, Shenzhen 518055, P. R. China.

[2] Shenyang National Laboratory for Materials Science, Institute of Metal Research, Chinese Academy of Sciences, Shenyang 110016, P. R. China.

[3] School of Materials Science and Engineering, University of Science and Technology of China, Shenyang 110016, P. R. China.

[4] State Key Laboratory of Geological Processes and Mineral Resources, China University of Geosciences, Beijing 100083, P. R. China.

[5] Faculty of Materials Science and Engineering / Institute of Technology for Carbon Neutrality, Shenzhen Institute of Advanced Technology, CAS, Shenzhen, 518055, P. R. China.

[6] Department of Physics and Astronomy, University of Manchester, Manchester M13 9PL, U.K.

[7] Advanced Technology Institute, University of Surrey, Guildford GU2 7XH, U.K.

\* Corresponding authors: bf.ding@siat.ac.cn (B.D.); bilu.liu@sz.tsinghua.edu.cn (B.L.L.)

## Abstract

**Optical device is a key component in our lives and organic liquid crystals are nowadays widely used to reduce human imprint. However, this technology still suffers from relatively high costs, toxicity and other environmental impacts, and cannot fully meet the demand of future sustainable society. Here we describe an alternative approach to colour-tuneable optical devices, which is based on sustainable inorganic liquid crystals derived from two-dimensional mineral materials abundant in nature. The prototypical two-dimensional mineral of vermiculite is massively produced by a green method, possessing size-to-thickness ratios of >$10^3$, in-plane magnetisation of >10 emu $g^{-1}$, and an optical bandgap of >3 eV. These characteristics endow two-dimensional vermiculite with sensitive magneto-birefringence response, which is several orders of magnitude larger than organic counterparts, as well as capability of broad-spectrum modulation. Our finding consequently permits the fabrication of various chromic devices with low or even zero-energy consumption, which can be used for sustainable optics.**

## Introduction

Sustainable global development requires the use of abundant natural materials, green processing technologies and energy-efficient operation in our future society, thus many technologies are undergoing profound changes.[1-4] Optical devices ranging from sub-micron-scale display pixels to meter-scale smart windows are important in modern society.[5-9] On the other hand, the current technologies face new challenges because of their environmental impact caused by the production of synthetic active materials, large energy consumption and complicated device architectures. Taken liquid crystal displays as an example, organic materials are most-commonly used as active materials in electro-optical devices but the relatively high toxicity and poor environmental stability of organic liquid crystals cause notable concerns.[10-13] In addition, current liquid crystal devices rely on the use of extremely-uniform microspheres as spacers and alignment layers, which are produced by repetitive rubbing of polyimide-coated glass substrates,[14] making the device production fairly complicated and resource consuming. Also, the existing electro-optical devices continuously consume energy to maintain working or stand-by states, regardless of whether the design is based on the vertically aligned or twisted nematic mode. Considering the huge size-to-thickness aspect ratio of two-dimensional (2D) materials and their resulting magnetic/electric/optical anisotropy, negligible energy consumption of magneto-optical devices, if driven by permanent magnets,[15-18] as well as abundant natural reserves of layered minerals suitable for production of 2D materials, we propose magneto-optical devices based on such minerals, which promises a way to realize sustainable optical technologies. Here we use 2D vermiculite (VMT), one of many micas, to prove the above concepts on sustainable optics.

**Results**

**Figure 1**a illustrates an eco-friendly method that can produce 2D minerals into large quantities with zero waste. As an instance, we employ it to produce monolayer VMT using mild operational conditions and only water and common salts like NaCl and LiCl as input chemicals (see Experimental Section for more details). VMT is a nontoxic (pharmaceutically acceptable) natural material with 600 million tons of reserves worldwide and effectively made of alternating stacked MgO octahedrons and SiO tetrahedrons,[19-21] as shown in Figure 1b. Its interlayer cations allow efficient ion exchange and exfoliation of the bulk materials (Figure 1c) into micrometre-sized 2D crystals of monolayer thickness. Moreover, the Fe-containing feature in centred octahedrons enable its sensitive magnetic response. Figure 1d shows the resulting 2D VMT colloidal dispersion in water, obtained on liter scale. The dispersion exhibits a pronounced Tyndall effect (Figure S1a). A zeta potential of -37.4 mV indicates the negative charge of the suspended 2D VMT, which supports the dispersion stability (Figure S1b). Atomic force microscopy characterisation reveals 2D crystals with smooth surfaces (Figure 1e) and an average lateral size of ~1.3 μm (Figure 1f) with a characteristic height of ~1.1 nm (Figure 1g), which corresponds to monolayer VMT

and the aspect ratio of above $10^3$. Transmission electron microscopy characterisation has confirmed that bulk VMT is exfoliated into atomically thin nanosheets without noticeable wrinkles and cracks (Figure 1h). The selected area electron diffraction demonstrates high crystallinity of the obtained 2D VMT (Figure 1h, inset). The observed in-plane distances for 2D VMT are shown in Table S1 and consistent with the atomic structure expected for monoclinic alluminosilicates.[22, 23]

Energy-dispersive X-ray spectroscopy maps of 2D VMT show the presence of Si, Al, Mg and Fe (Figure 1i). We use electron probe microanalysis to quantify four main components in VMT, namely $SiO_2$, $MgO$, $Al_2O_3$ and FeOT, and their content is found to be 39.35, 24.42, 10.99 and 6.63 wt%, respectively (Table S2). Here FeOT refers to the total Fe content in the form of various ferrous compounds. Magnetic measurements reveal that 2D VMT possesses a notable anisotropy between in-plane and out-of-plane magnetic susceptibilities that start saturating above a few Tesla at liquid-helium temperatures, with the in-plane easy magnetisation axis (Figure 1, j and k). We have also prepared 2D VMT laminate films by filtration and observed their rotation such that the films' surface aligned along magnetic field (Figure S2, a and b, and Video S1), in agreement with the susceptibility measurements. Optical anisotropy provides *in situ* evidence for magnetic alignment. The light intensity was found to be stronger when observed in the direction perpendicular to the applied magnetic field than parallel to it (Figure S2, c to f), which is a scattering result for 2D crystals being parallel to magnetic field. At the same time, optical absorption measurements indicate rather high transparency of the 2D VMT suspensions (Figure 1l, inset and Figure S1c). Using the Tauc plot [$(Ahv)^{\frac{1}{2}}$ as a function of the photon energy $hv$, where $A$ is the absorbance], we estimate the band gap to be ~3.9 eV, which corresponds to the absorption edge at ~320 nm (Figure 1l) and transparency of 2D VMT for the whole visible spectrum.

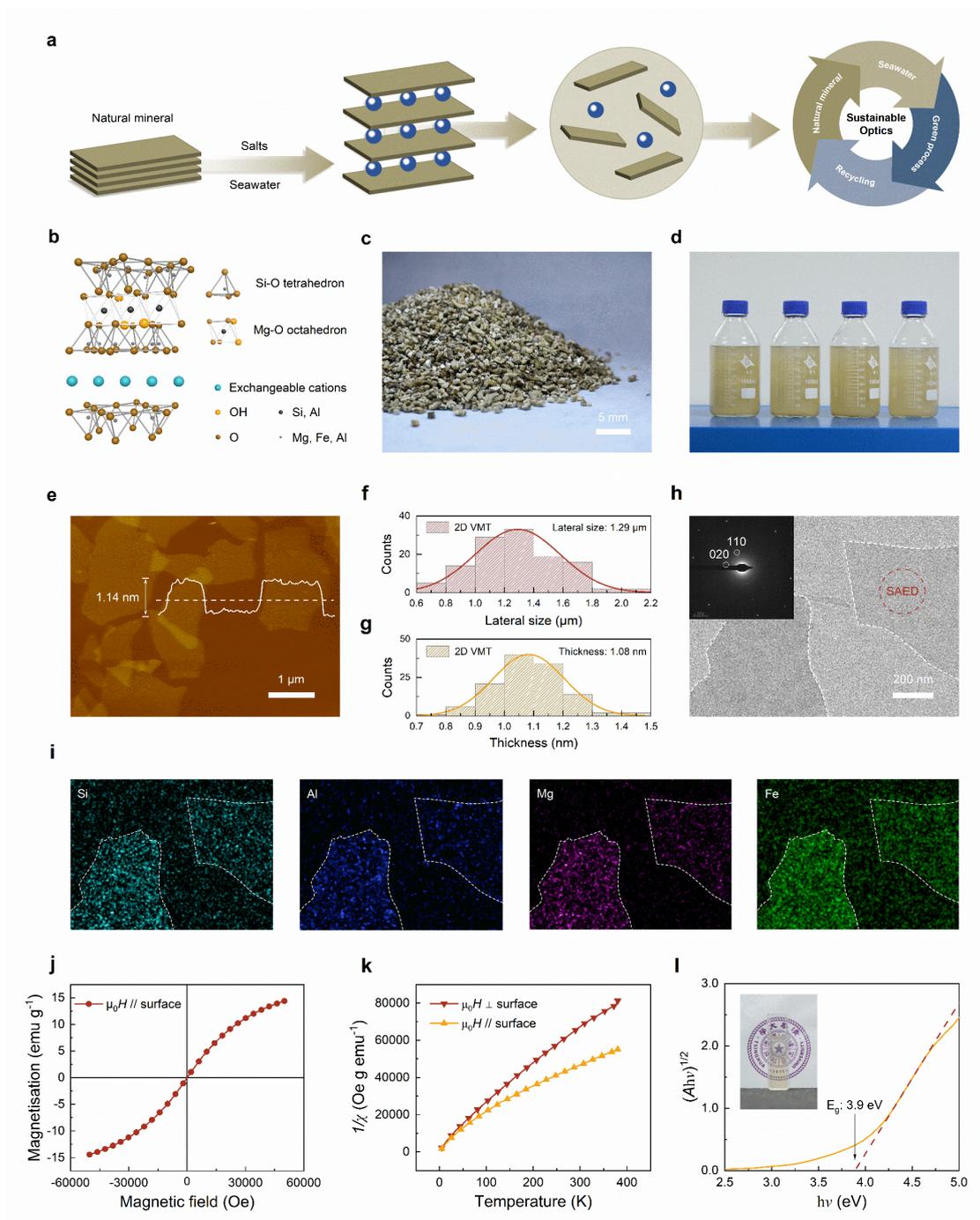

**Figure 1. Synthesis and characterisation of 2D VMT.** a) Schematic of an eco-friendly method that can produce 2D minerals into large quantities. b) Atomic structure of 2D VMT: VMT monolayers contain an octahedral MgO layer sandwiched by a pair of SiO tetrahedral layers. c) Photo of typical bulk VMT available as a mineral. d) 2D VMT dispersions prepared in liter quantities. e) Atomic force microscopy image of 2D VMT deposited on a substrate. Statistics for the crystals' lateral size (f) and thickness (g). h) Typical transmission electron micrograph and selected area diffraction pattern (inset)

of monolayer VMT. i) Energy-dispersive X-ray maps for Si, Al, Mg and Fe. j) In-plane magnetisation for a 2D VMT laminate at 5 K. k) Its in-plane and out-of-plane inverse magnetic susceptibilities as a function of temperature at 0.1 Tesla. l) Tauc plot for 2D VMT. Inset: Optical image of 2D VMT aqueous dispersion in a 10 mm thick cuvette illustrating transparency of the VMT suspension.

If 2D VMT suspensions are shaken and placed between crossed polarisers, they can exhibit clear birefringent patterns (**Figure 2**, a and b). As the concentration of 2D VMT increases, we observe a sequence of isotropic, biphasic and nematic phases (Figure 2c). Figure 2d shows the fraction of the nematic phase as a function of the VMT concentration, which is estimated by calculating the area exhibiting birefringence.[6] Isotropic-to-nematic phase transition is found to occur at concentrations between 0.32 and 0.56 vol%. These observations show that dilute aqueous dispersions of 2D VMT can behave as inorganic liquid crystals without need of any organic components.

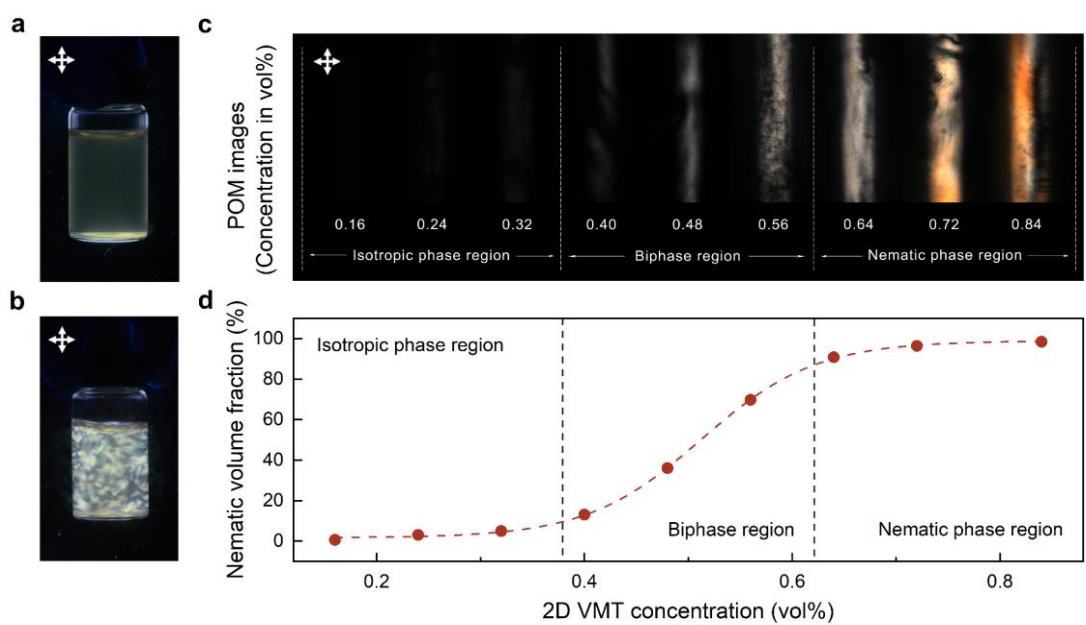

**Figure 2. 2D VMT as an inorganic liquid crystal.** a) Optical image of a 2D VMT dispersion in the steady state. b) Flow-induced birefringence after shaking. Crossed polarisers are used in both (a and b) cases. c) Optical images of capillary tubes (1.0 mm in diameter) containing VMT dispersions in different concentrations. The dispersions with concentrations of 0.64, 0.72 and 0.84 vol% show clear birefringent regions,

indicating a nematic phase. Dispersions with concentrations below 0.32 vol% show no birefringence and are isotropic. The 0.40, 0.48 and 0.56 vol% dispersions in between are in the biphasic state. d) Nematic fraction as a function of VMT concentration.

Considering the observed shape and magnetic anisotropy, we examine the magneto-optical response of isotropic 2D VMT suspensions. Their phase-retardation capability is studied first. Birefringence $\Delta n$ give rise to the phase retardation $\delta = \frac{2\pi \Delta n L}{\lambda}$ where $L$ is the optical path and $\lambda$ is the wavelength. **Figure 3**a shows the magneto-birefringence effect of our VMT suspensions in magnetic fields from 0 to 0.8 Tesla. In small fields, the magneto-birefringence follows the Cotton-Mouton law $\Delta n = \lambda C (\mu_0 H)^2$ evolving proportionally to the square of the field strength $\mu_0 H$. The Cotton-Mouton coefficient $C$ describes the sensitivity of magneto-optical response and, for concentration of 0.16 vol%, reaches a value of ~1,080 m$^{-1}$T$^{-2}$ (Figure 3b), which is orders of magnitude larger than any other transparent birefringent medium (Figure 3c).[24] The saturation birefringence $\Delta n_{sat}$ of a 0.16 vol% VMT dispersion, calculated by extrapolating measured birefringence curve to $\mu_0 H \to \infty$, is found to be ~3.8×10$^{-5}$ at the wavelength of 650 nm (Figure S3). This corresponds to a phase retardation of about 1.2π for the optical distance $L$ of 10 mm. Combing the saturation birefringence of 2D VMT suspensions with other concentration, specific birefringence $\Delta n^p$ is estimated to be 5.45×10$^{-2}$ according to the linear fit on $\frac{\Delta n_{sat}}{S_{sat}} = \Delta n^p \varphi$ (Figure 3d), where we assume that the saturation order parameter $S_{sat} = -0.5$ at $\mu_0 H \to \infty$ for an ideal collinearly aligned system.[25] The $\Delta n^p$ represents the maximum strength of magneto-optical response for volume fraction $\varphi = 1$ within a fully ordered system.[26]

The magneto-transmittance maps in Figure 3 show that the phase-retardation capability of VMT suspensions allows light modulation over the whole visible spectrum. Moreover, the light modulation is highly controllable and can be adjusted by changing VMT concentration (Figure S4) and optical path (Figure S5) so that the second interference maxima can easily be reached (see the second bright strip in Figure 3e).

Importantly, the observed light modulation and its dependence on wavelength and magnetic field are quantitatively described by theory,[24-26] which not only allows accurate simulation of the experiment (see Experimental Section for more details, Figure 3f and Figure S6) but also provides a way to design magneto-optical modulators with tuneable characteristics.

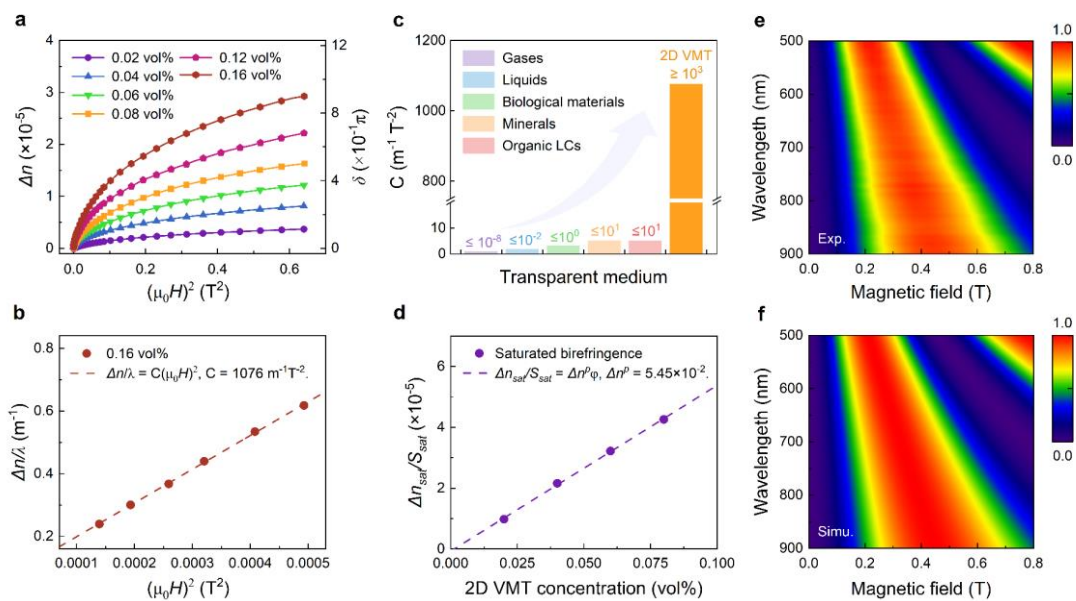

**Figure 3. Magneto-optical response of 2D VMT suspensions.** a) Magneto-birefringence for different concentrations of 2D VMT in fields up to 0.8 Tesla. b) Cotton-Mouton coefficient $C$ for VMT dispersion with a concentration of 0.16 vol%. c) Comparison of the measured coefficient with those of gases, liquids, biological materials, minerals and organic liquid crystals. d) Saturation birefringence of 2D VMT dispersions. Data in (a to d) are collected under 650 nm laser irradiation. (e and f), Experimental and simulated maps, respectively, of transmitted light intensity for a 0.24 vol% dispersion in a cuvette with $L$ = 16 mm. The colour bars in (e) and (f) represent the relative light intensity.

The described characteristics of 2D VMT suspensions allow us to design and demonstrate various prototype magneto-optical devices (**Figure 4**a). First, we demonstrate an optical switch between dark (off) and bright (on) states with the dynamic control of magnetic field (Figure S7, a and b). Permanent magnets were used to operate this device, indicating a possibility of zero energy consumption during device operation. We have also measured time response of such switchers by using pulsed magnetic fields reaching 1.2 Tesla within 2 μs. By fitting the observed optical intensity $I$ as a function of time $t$ and assuming the exponential decay, $I \propto e^{-\frac{t}{\tau}}$, we find the characteristic time $\tau \approx 40$ ms, demonstrating surprisingly rapid response for such magneto-optical switches (Figure S7c). We also find little difference in $I$ if increasing or decreasing magnetic field (Figure S7d). Furthermore, after 150 cycles, light intensity for the on-state remained at 98% of that in the first cycle and the final on-off ratio exceeded 300 (Figure S7e), which shows good durability of the magneto-optical switchers.

The device principle can also be used for making see-through displays (Figure 4). For example, Figure 4b shows a colour sequence of black, yellow, orange, red, purple, blue and green which appear with increasing magnetic field, covering practically the full visible light spectrum. The individual colours can be presented on the Commission Internationale de l'Eclairage chart, and Figure 4c shows their colour evolution from 0.2 to 2.0 Tesla. The controllable choice of colour codes allows designing of rainbow patterns. This is illustrated in Figure 4d that shows letters THU and SGC under static fields of 1 and 2 Tesla. Furthermore, with reference to chameleons changing their skin colours, we mimic such changes by demonstrating a smart "chameleon" in Figure 4d. Our chameleon changes its colours in response to its own environment, that is, the magnetic field. Once the external stimulus is removed, the chameleon disappears, and the colours also fade if the crossed polarisers are removed (Figure 4d and Video S2). The demonstrated coloured patterns are achieved using an exceptionally simple device architecture, without using spacers, transparent electrodes or alignment layers, which can make applications of such magneto-optical devices commercially competitive in colour displays, smart sensors, digital tags and security labels.

To strengthen the above point, we also demonstrate a visual force indicator that is based on a hydrogel containing 2D VMT and polymerised under 0.8 Tesla (see the "Magneto-optical devices" section). The hydrogel exhibits strong birefringence that depends on mechanical strain (Figure 4e) and can withstand a stress of >600 kPa and elastic compression up to 80% (Figure 4f). The indicator shows first-order pure white and bright yellow under 20% compression, followed by the emergence of abundant second-order interference colours in sequence for deformation from 20% to 50% (Figure 4f and Video S3).

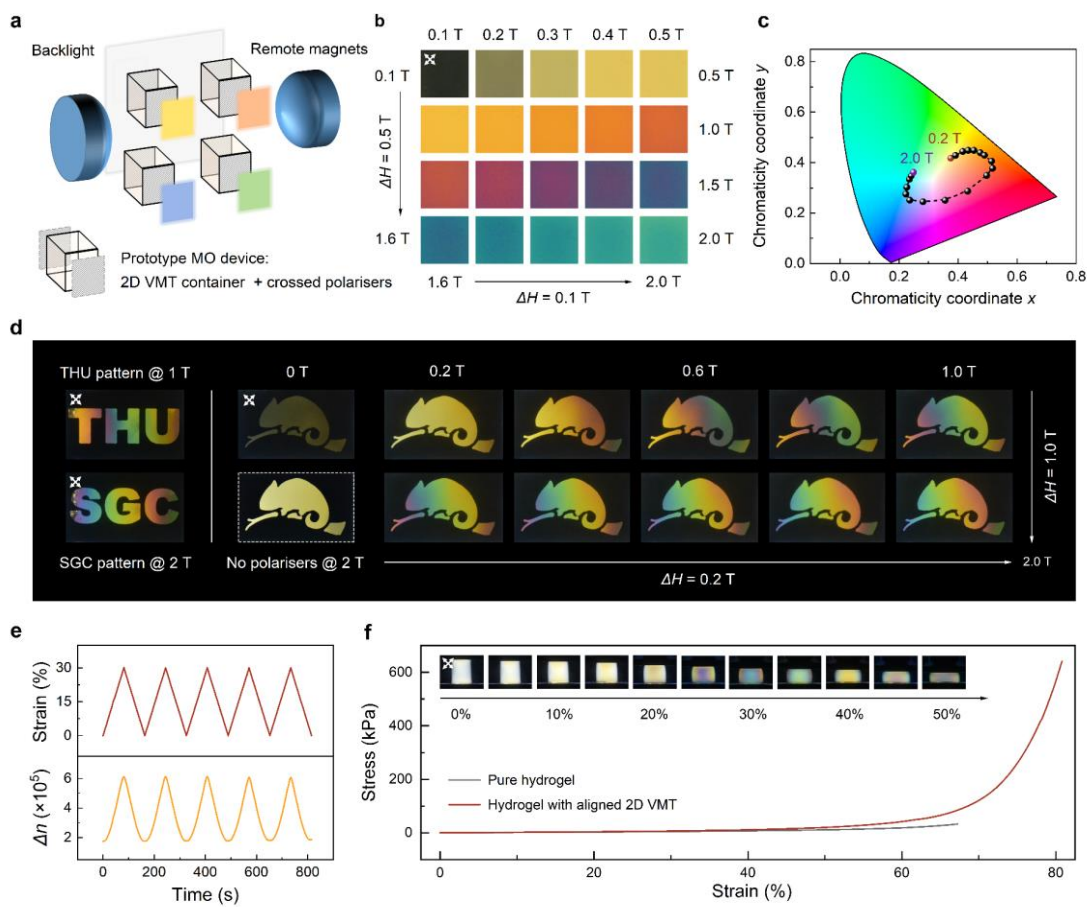

**Figure 4. Prototype magneto-optical devices.** a) Their schematic. b) Interference colours observed for a 0.24 vol% VMT dispersion in fields between 0.1 and 2.0 Tesla. c) Evolution of colours in (b) with increasing the magnetic field, presented as coordinates in the standard Commission Internationale de l'Eclairage 1931 chart. d) See-through patterns showing letters THU and SGC with gradient colours at different magnetic fields and a "smart chameleon" changing its colours in response to the magnetic stimulus. The gradient colours within each image appear due to the use of a slanted cuvette that provides a horizontal gradient of the optical path $L$ (see the "Magneto-optical devices" section). e) Changes in birefringence $\Delta n$ with strain for our force indicator based on a hydrogel containing 2D VMT. f) Strain-stress curve for the indicator. Inset: Its colour changes as seen in the direction perpendicular to applied stress.

Finally, let us note that the 2D VMT used in this study can be prepared in a low cost and green manner. First, a closed-loop exfoliation can be achieved by collecting and reusing all chemicals and the remaining unexfoliated VMT, offering zero waste within the whole preparation process (Figure S8a). Second, we also used simulated seawater to prepare 2D VMT dispersions that showed the same optical behaviour with those prepared by using deionized water and NaCl chemicals (see Experimental Section for more details, Figure S8, b and c). This offers the use of natural seawater as dispersants for industrial scale manufacturing of 2D-VMT-based devices. Third, beyond the device lifecycle, the used mineral suspension can be returned into nature as a fertilizer, soil conditioner or seed germination medium. Therefore, the demonstrated approach to magneto-optical devices meets all expectations from the sustainability perspective.

In summary, we have explored a possibility of magneto-optical devices based on 2D mineral vermiculite, which feature naturally abundance, low cost, eco-friendly production and recyclability as well as energy-efficient device operation. Considering many types of 2D minerals with different properties, their functional uses should be further exploited in search for novel opportunities and advanced technological applications.

## Experimental Section

**Production of monolayer 2D VMT.** Two different kinds of VMT (sizes: 2-3 mm, Sigma-Aldrich, USA; sizes: 1.4-4 mm, Xinjiang Yuli Xinlong Vermiculite Co., Ltd., China), sodium chloride (purity >99.5%, Shanghai Macklin Biochemical Co., Ltd., China), and lithium chloride (purity >99.0%, Shanghai Macklin Biochemical Co., Ltd., China) were used as received. We used an ion-exchange method to exfoliate bulk VMT to prepare 2D VMT[20, 21], but some simplifications were made to avoid unnecessary use of extra equipment (e.g., reflux), chemicals (e.g., $H_2O_2$) and aggressive conditions such as high temperatures. In our method, the bulk VMT was soaked in a saturated NaCl solution, and stirred at 80 °C for 36 h, followed by repeated washing with deionized water. The process was repeated with the NaCl solution being changed to a 2M LiCl solution. 2D VMT was obtained after 5 minutes of bath sonication. The unexfoliated VMT, NaCl and LiCl solution were collected and reused in the next run of the above process. Saline with a NaCl mass fraction of 35‰ was used as simulated seawater in some experiments.

**Material characterisation.** Atomic force microscopy (Tapping mode, Cyper ES, Oxford Instruments, USA) and transmission electron microscopy (200 kV, F200X Talos, FEI, USA) were used to characterise morphology of exfoliated VMT. The dispersion stability was studied by a zeta potential analyser (Zetasizer Nano-ZS90, Malvern, UK). Magnetisation of VMT laminates was measured as a function of temperature and magnetic field at a magnetic properties measurement system (MPMS-7 XL, Quantum Design, USA) for both in-plane and out-of-plane geometries. The magnetic field was either parallel or perpendicular to the surface of the VMT films. The optical measurements were based on using a UV-vis absorption spectrophotometer (UV-2600, Shimadzu, Japan). A polarised optical microscope (Imager A2m, Carl Zeiss, Germany) was used to take polarised optical images of 2D-VMT-containing capillaries.

**Magneto-optical measurements.** A quartz cuvette with a 10 mm optical path was used as a container of the 2D VMT dispersion and was placed in the centre between two poles of a magnet. Orthogonal polarisers (GL10-A, Thorlabs Inc., USA) were placed in the optical path perpendicular to the magnetic field direction, and the angle between them was 45 degrees. The intensity of transmitted light was recorded by a power meter (PM200, Thorlabs Inc., USA). The polarization of transmitted light was detected by a polarimeter (PAX1000, Thorlabs Inc., USA).

For the transmittance maps, we used a tungsten lamp to provide incident white light, and the transmitted signals were collected by a spectrometer (QE65 Pro, Ocean Optics, China). Unless stated otherwise, for quantitative characterisation of magneto-optical we used a 650 nm laser.

**Theoretical simulations.** Our simulations are based on two independent variations of birefringence to wavelength and magnetic field, respectively. First, at a given laser wavelength $\lambda$, the birefringence $\Delta n$ (phase retardation $\delta$) is proportional to an orientational order parameter $S$ given by generalized second order Langevin function, which includes the influence of magnetic field $\mu_0 H$. It follows $\delta = \frac{2\pi \Delta n L}{\lambda}$, $\Delta n = \Delta n^p \varphi S$, $S = \frac{\int_1^{-1} \frac{1}{2} e^{Ax^2}(3x^2-1)dx}{\int_1^{-1} e^{Ax^2} dx}$, and $A = \frac{\Delta \chi}{2k_B T}(\mu_0 H)^2$, where $L$ is the optical path, $\Delta n^p$ the specific birefringence, $\varphi$ the volume fraction, $\Delta \chi$ the susceptibility anisotropy, $k_B$ the Boltzmann constant, and $T$ the temperature. Second, at a given magnetic field $\mu_0 H$, the birefringence dispersion relation with laser wavelength $\lambda$ can be found using the expression $\delta = k' \frac{\lambda \lambda_*^2}{\lambda^2 - \lambda_*^2}$, where $k'$ is a numerical constant and $\lambda_*$ the resonant wavelength. Combining the above dependences obtained by fitting the experimental curves, we construct a theoretical transmittance map in Figure 3f.

**Magneto-optical devices.** The devices are based on essentially the same setup as described in 'Magneto-optical measurements'. The on-off switch and dynamic monochromic display used cuvettes with an optical path of 10 mm and 16 mm, respectively. Accurate colours in Figure 4b were taken by a Nikon D7000, AF-S DX Nikkor 18-140 mm f/3.5-5.6G ED VR @140 nm with ISO 800, F8, 1/250. The THU, SGC and chameleon images were made using patterns cut in black cardboard and the cuvette had a gradient depth of 8 to 16 mm. The force indicator was a hydrogel cylinder with an initial diameter of 14 mm and a height of 18 mm. The hydrogel was made as follows. Acrylamide (purity >99.9%, Aladdin, USA), poly(ethyleneglycol) diacrylate (PEGDA, average Mn = 700, Sigma-Aldrich, USA), 2-Hydroxy-4′-(2-hydroxyethoxy)-2-methylpropiophenone (Irgacure 2959, Aladdin, USA), and (3-Acrylamidopropyl) trimethylammonium chloride (74%-76% in water, Shanghai Macklin Biochemical Co., Ltd., China) were mixed with a 0.08 vol% VMT dispersion and the mixture was stirred for 15 minutes to ensure its homogeneity. The mixture was then placed in a glass container and exposed to a constant field of 0.8 Tesla. Polymerisation was induced by exposure to UV light (365 nm at 50 W) for 5 minutes. The free-standing force indicator was extracted from the container after the polymerisation process. All the chemicals were used as received. The strain-stress curves were collected by a universal testing system (5900 Series, Illinois Tool Works Inc., USA) and the changing colours were recorded during testing.

## Supporting Information

Supporting Information is available.

## Acknowledgements

We acknowledge support by the National Natural Science Foundation of China (Nos. 51920105002, 52125309, 51991343, and 51991340), the Guangdong Innovative and Entrepreneurial Research Team Program (No. 2017ZT07C341), the Shenzhen Basic Research Project (Nos. WDZC20200819095319002 and JCYJ20190809180605522), the National Key R&D Program (2018YFA0307300), and the Bureau of Industry and Information Technology of Shenzhen for the "2017 Graphene Manufacturing Innovation Centre Project" (No. 201901171523). We also acknowledge Liusi Yang, Yikun Pan, Pengyuan Zeng, Hao Xu, Youan Xu for fruitful discussions or manuscript revision.

## Conflict of Interest

The authors declare no conflict of interest.